\documentclass{article}

\usepackage{arxiv}

\usepackage[utf8]{inputenc} 
\usepackage[T1]{fontenc}    
\usepackage{hyperref}       
\usepackage{url}            
\usepackage{booktabs}       
\usepackage{amsfonts}       
\usepackage{nicefrac}       
\usepackage{microtype}      
\usepackage{lipsum}
\usepackage{graphicx}

\title{A comparison of Vietnamese Statistical Parametric Speech Synthesis Systems}

\author{
  Huy Kinh Phan\thanks{The first two authors contributed equally} \\
  Zalo Group, VNG Corporation\\
  Hanoi, Vietnam \\
  \texttt{phanhuykinh@gmail.com} \\
   \And
  Viet Lam Phung \\
  Zalo Group, VNG Corporation\\
  Hanoi, Vietnam \\
  \texttt{lam.phungviet33@gmail.com} \\
   \And
  Tuan Anh Dinh \\
  Zalo Group, VNG Corporation\\
  Hanoi, Vietnam \\
  \texttt{tuanad121@gmail.com} \\
   \And
  Bao Quoc Nguyen \\
  Zalo Group, VNG Corporation\\
  Hanoi, Vietnam \\
  \texttt{baonq6@vng.com.vn} \\
  Information and Communication Technology University, Thai Nguyen University \\
  Thai Nguyen, Vietnam \\
}

\begin{document}
\maketitle

\begin{abstract}
In recent years, statistical parametric speech synthesis (SPSS) systems have been widely utilized in many interactive speech-based systems (e.g.~Amazon's Alexa, Bose's headphones). To select a suitable SPSS system, both speech quality and performance efficiency (e.g.~decoding time) must be taken into account. In the paper, we compared four popular Vietnamese SPSS techniques using: 1) hidden Markov models (HMM), 2) deep neural networks (DNN), 3) generative adversarial networks (GAN), and 4) end-to-end (E2E) architectures, which consists of Tacontron~2 and WaveGlow vocoder in terms of speech quality and performance efficiency. We showed that the E2E systems accomplished the best quality, but required the power of GPU to achieve real-time performance. We also showed that the HMM-based system had inferior speech quality, but it was the most efficient system. Surprisingly, the E2E systems were more efficient than the DNN and GAN in inference on GPU. Surprisingly, the GAN-based system did not outperform the DNN in term of quality.
\end{abstract}

\keywords{Hidden Markov models \and Deep neural networks \and Statistical parametric speech synthesis \and End-to-end}

\section{Introduction}

\begin{figure}[t]
\centering
\includegraphics[width=0.9\textwidth]{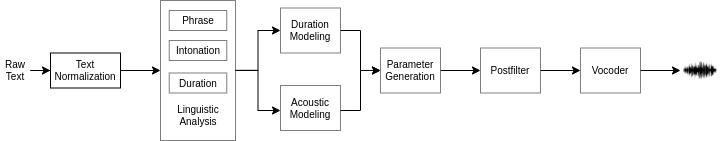}
\caption{General flowchart of SPSS systems}\label{fig:flowchart}
\end{figure}

A general flowchart of a SPSS system is shown in Figure~\ref{fig:flowchart}. Given input text, a text normalization component detects and expands non-standard words such as abbreviations, numbers, date, time ... to obtain normalized text~\cite{SPROAT2001287}. The linguistic features such as phonemes, phoneme context, word boundaries, phrase breaks, ... are extracted from the normalized text by a linguistic analysis component~\cite{tokuda}. An acoustic modelling component obtains the statistics of the acoustic features (e.g.~fundamental frequency (F0), mel-generalized cepstrum coefficients (MCC)~\cite{To94mel-generalizedcepstral}, and band-aperiodicity~\cite{MORISE201657}) of each phoneme from its linguistic features using decision tree regression in combination with HMM~\cite{tokuda} or deep neural networks (DNN)~\cite{Zen13} or conditional generative adversarial models (cGAN)~\cite{ganTTS}. It was showed that the cGAN-based acoustic modelling outperformed the DNN-based acoustic modelling because cGAN could address over-smoothing problem~\cite{ganTTS}. Similarly, a duration modelling component predicts the duration of each phoneme from the linguistic features. A parameter generation component produces the most likely trajectory of acoustic features from the obtained statistics and phoneme duration using a maximum likelihood criterion~\cite{Tokuda95}. The generated acoustic features can be limited by the over-smoothing problem; therefore, a postfiltering component can be utilized to address the problem using global variance scaling~\cite{Hana12} or modulation spectrum scaling~\cite{Takamichi}. Another postfiltering approach is to use voice conversion techniques such as DNN~\cite{dnnPF} or GAN~\cite{ganPF} to map synthesized speech to natural speech. Finally, a vocoder is used to generate synthesized speech from the filtered acoustic features (e.g.~WORLD vocoder~\cite{morise16}). A comparison of STRAIGHT, glottal sinusoidal vocoding was conducted; indicating that waveform generation method of a vocoder is essential for quality improvement of SPSS systems. 

However, some components are heavily knowledge-driven such as linguistic and acoustic analysis; which can be suboptimal. In an end-to-end (E2E) architecture, the five components: linguistic analysis, acoustic modeling, duration modelling, parameter generation, and postfiltering are replaced by encoder-attention-decoder networks~\cite{Tacotron17, transformertts} to be able to effectively optimize the mapping from input text to acoustic features (e.g.~80-dimensional mel-spectrum). At last, a neural vocoder such as WaveNet produced a synthesized waveform from the generated mel-spectrogram~\cite{wavenet}.
By leveraging the encoder-attention-decoder networks in combination with neural vocoders, the E2E SPSS systems can address the over-smoothing problem as well as achieving the state-of-the art quality~\cite{Tacotron17}. But the high quality comes with a price of high computational complexity as well as huge models; therefore, the E2E systems highly depend on GPU resources to achieve a real-time performance. In contrast, a HMM-based SPSS system can perform on a single CPU, and a DNN-based SPSS system can run on multiple CPUs. To improve the performance of the E2E systems, efficient neural vocoders were proposed using flow (known as WaveGlow)~\cite{glow} or linear prediction (known as LPCNet)~\cite{LPCNet}. 

Researchers have attempted to build high quality Vietnamese SPSS systems in the last two decades. A data processing scheme proved its efficacy in optimizing naturalness of E2E SPSS systems~\cite{Lam20} trained on Vietnamese found data. Text normalization methods were explored; utilizing regular expressions and language model~\cite{textnorm} or sequence-to-sequence models~\cite{textnorm2}. New prosodic features (e.g.~phrase breaks) were investigated and showed their efficacy in improving naturalness of Vietnamese HMM-based SPSS systems~\cite{prosody13, prosody13b}. Different types of acoustic models were investigated such as HMM~\cite{Vu09,Glottal15} or DNN~\cite{dnnTTS}. In postfilterin, it was shown that a global variance scaling method may destroy the tonal information~\cite{postrocess2, postrocess}; therefore, exemplar-based voice conversion methods were utilized in postfiltering to preserve the tonal information~\cite{postrocess, Nguyen2017}. Moreover, a Vietnamese Language and Speech Processing (VLSP) evaluation is organized annually where the progress of developing Vietnamese SPSS systems is evaluated~\cite{vlsp}. Our E2E SPSS system achieved the best performance of VLSP 2019~\cite{Lam20}. Our E2E SPSS system consits of Tacontron~2~\cite{tacotron2} and WaveGlow vocoder~\cite{glow}.

In the paper, we first described four Vietnamese SPSS systems: 1) HMM-based SPSS system, 2) DNN-based SPSS system, 3) cGAN-based SPSS system, 4) and E2E SPSS system in Section~\ref{sec:Vietnamese_SPSS}. In Secton~\ref{evaluation}, we compared the four systems in term of speech quality and efficiency.

\section{Vietnamese SPSS systems} \label{sec:Vietnamese_SPSS}
In the section, we described the structures of four Vietnamese SPSS systems.

\subsection{HMM-based SPSS system}\label{sec:hmm}
\begin{figure}[t]
\centering
\includegraphics[width=0.7\textwidth]{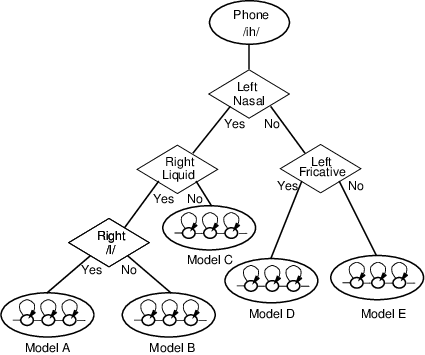}
\caption[width=0.8\columnwidth]{A linguistic decision tree \cite{decisiontree}}\label{fig:tree}
\end{figure}

We utilized the HTS framework~\cite{hts-link} to build our Vietnamese HMM-based SPSS system. In the training process, we prepared audio files and corresponding labels. The labels contain full-context labels of each phoneme in input text and phoneme time-stamps in a unit of 100~$\mu$s. The time-stamps are used to calculate the phoneme durations. Acoustic features are frame-based F0, MCC, and band-aperiodicity and their corresponding deltas and delta-deltas. In other words, we not only model the static acoustic features but also their dynamic in time. The static and dynamic features are combined to obtain synthesized F0, MCC, and band-aperiodicity. A WORLD vocoder is used to generate an waveform from the acoustic features~\cite{morise16}. For acoustic modelling of each phoneme, binary decision trees are built from linguistic features and a linguistic question set for clustering acoustic spaces corresponding to F0, MCC, and band-aperiodicity, and their deltas and delta-deltas~\cite{decisiontree}. An example of a question in the question set is "is current phoneme ih?" (as shown in Figure~\ref{fig:tree}). For duration modelling, another binary decision tree is built from linguistic feature and another linguistic question set for clustering the phoneme duration space. In total, there are two linguistic question sets for building decision trees. For each cluster, a three-state (excluding starting and ending states) HMM is built to capture the statistics (e.g.~mean and standard deviation of Gaussian observation probability) of a context-dependent phoneme (as shown in Figure~\ref{fig:tree})~\cite{tokuda}. In inference process, full-context labels are extracted from input text by a linguistic analysis component. Given the phoneme labels, suitable context-dependent HMMs are selected using the binary decision trees. A parameter generation algorithm uses the sequence of HMMs to generate smooth acoustic feature trajectories. A modulation spectrum-based postfiltering is used to compensate for the over-smoothing problem~\cite{Takamichi}. Finally, an waveform is synthesized from the filtered acoustic features using the WORLD vocoder.

In the Vietnamese HMM-based SPSS system, we used the same full-context labels in linguistic analysis as in~\cite{Vu09, prosody13}. In acoustic modelling, the two question sets for duration and acoustic modelling were also the same as in~\cite{Vu09, prosody13}. For acoustic features,  we used MCC of the full-band aperiodicity instead of the band-aperiodicity of WORLD vocoder. Our unofficial experiments showed that the former has better performance than the latter. We defied a comprehensive set of regular expressions for text normalizaion~\cite{textnorm}. 

\subsection{DNN-based SPSS system} \label{sec:dnn}

\begin{figure}[t]
\centering
\includegraphics[width=0.5\textwidth]{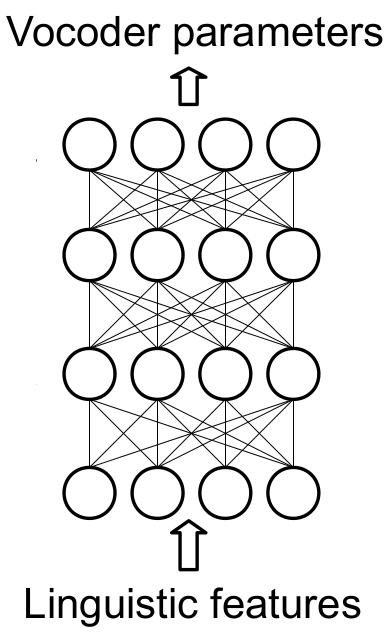}
\caption[width=0.5\columnwidth]{DNN-based mapping from frame-based linguistic features to frame-based vocoder/acoustic features}\label{fig:dnn-mapping}
\end{figure}

We used the Merlin framework~\cite{merlin1fba9fc9b9884b24ae648bfc6c636fba} to build our Vietnamese DNN-based SPSS system. The DNN-based SPSS system address the over-smoothing problem by replacing the decision trees with DNN models. There are also two linguistic question sets in acoustic and duration modelling. The input vectors of the DNN consists of the answers for the two question sets. The output vectors of the DNN are also F0, MCC, band-aperiodicity, and their deltas and delta-deltas. We can think about the SPSS system as a direct mapping from linguistic features to acoustic features using DNN (as shown in Figure~\ref{fig:dnn-mapping}). The DNN can represent the complex. non-linear relationship better than the decision trees~\cite{dnnTTS}. A simple postfiltering method was used to emphasize the formants of synthesized speech~\cite{simplePF}. The modulation spectrum-based postfiltering was not used. The WORLD vocoder was utilized for waveform generation.

In the Vietnamese DNN-based SPSS system, the full-context labels and the linguistic decision sets are the same as in Section~\ref{sec:hmm}. The acoustic features were also the same as in Section~\ref{sec:hmm}.  

\subsection{cGAN-based SPSS system}

Traditional GANs have a generative model or a generator ($G$) and a discriminative model or a discriminator ($D$). that together play a min-max game. Component $G$ tries to fool component $D$ by generating outputs similar to the real data, while component $D$ is trained to distinguish the output of component $G$ from real data.  Component $G$ is a mapping function from random noise z to $v$, $G:\left\{ z \right\} \rightarrow v$~\cite{fellow}. In contrast, a cGAN model learns a mapping from an input u and random noise z to v, $G:\left\{ u,z\right\} \rightarrow v$. The cGAN model has both $G$ and $D$ conditioned on input u~\cite{ganTTS}, trained with the objective function $\mathcal{L}(D,G)$: 
\begin{equation}
\min_{G}\max_{D}\mathcal{L}(D,G) = \mathbb{E}_{u,v}\left[\log D(u,v)\right]+\mathbb{E}_{u,z}\left[\log(1-D(u,G(u,z)))\right]
\end{equation}

In our cGAN, we did not use random noise z because it has proven ineffective for generator $G$~\cite{ganTTS}. Instead, our generator mapped linguistic features $u$ to acoustic features $v$ (similar to the DNN-based mappping in Section~\ref{sec:dnn}). The linguistic and acoustic feature are the same as in Section~\ref{sec:dnn}. In addition to the adversarial loss function $\mathcal{L}(D,G)$ in Equation~\ref{eq1}, we also minimized the mean squared error between the output of $G$ and the ground truth~\cite{ganTTS}. The WORLD vocoder was used to synthesize waveform from generated acoustic features. The cGan-based mapping may address the over-smoothing of generated features better than DNN-based mapping~\cite{ganTTS}. Our implementation was based on~\cite{gan-link}. 

\subsection{End-to-end SPSS system}

Our end-to-end TTS system have two components: 1) an encoder-attention-decoder-based acoustic model and 2) neural vocoder~\cite{Lam20}. The acoustic model converts a sequence of syllables with prosodic punctuations to a 80-dimensional mel-spectrogram. The neural vocoder generates speech from the 80-dimensional mel-spectrogram. In normalized text, we consider tokens, inserted prosodic punctuation as syllables. In the paper, we utilized Tacotron~2~\cite{tacotron2} for acoustic modeling, and WaveGlow vocoder~\cite{glow}. As a result, our end-to-end system can achieve a real-time inference speed on GPU. 

Generally, the acoustic model of an E2E SPSS system has a encoder-decoder structure~\cite{seq2seq} equipped with an attention mechanism~\cite{attention}. The encoder-decoder structure maps a source sequence $\mathbf{x}_{1:n}=\left( \mathbf{x}_1, ..., \mathbf{x}_n \right)$ to a target sequence $\mathbf{y}_{1:m}=\left( \mathbf{y}_1, ..., \mathbf{y}_m \right)$ which can have different lengths, i.e.~$m \neq n$. In our case, the source sequence $\mathbf{x}_{1:n}$ is the sequence of syllables represented as one-hot vectors. The target sequence $\mathbf{y}_{1:m}$ is the sequence of 80-dimensional mel-spectrum. First of all, the encoder (Enc) maps the source sequence $\mathbf{x}_{1:n}$ to a sequence of hidden representaitions $\mathbf{h}_{1:n}=\left( \mathbf{h}_1, ..., \mathbf{h}_n \right)$. The decoding of the target sequence is autoregressive, which means the previously generated vectors are considered as addition input at each decoding step $t$. To generate an output vector $\mathbf{y}_t$, a weighted sum of $\mathbf{h}_{1:n}$ forms a context vector $\mathbf{c}_t$, where the weight vector is a calculated attention probability vector $\mathbf{a}_t=\left( a_t^{(1)}, ..., a_t^{(n)} \right)$. We can think about the attention probability $a_t^{(k)}$ as the importance of hidden representation $h_k$ at time $t$. Finally, the decoder (Dec) use the context vector $\mathbf{c}_t$ and the previously generate features $\mathbf{y}_{1:t-1}$ to decode $\mathbf{y}_t$. Note that the attention calculation and decoding process take the previous hidden state of the decoder $\mathbf{q}_{t-1}$ as a input. We can formulate the above procedure as follows:
\begin{equation}
    \mathbf{h}_{1:n} = \textrm{Enc} \left( \mathbf{x}_{1:n} \right)
\end{equation}

\begin{equation}
    \mathbf{a}_t = \textrm{attention} \left( \mathbf{q}_{t-1}, \mathbf{h}_{1:n} \right)
\end{equation}

\begin{equation}
    \mathbf{c}_t = \sum_{k=1}^n a_t^{(k)} \mathbf{h}_k    
\end{equation}

\begin{equation}
    \mathbf{y}_t, \mathbf{q}_t = \textrm{Dec} \left( \mathbf{y}_{1:t-1}, \mathbf{q}_{t-1}, \mathbf{c}_t \right)
\end{equation}

We used Tacotron~2~\cite{tacotron2} for acoustic modelling. Our implementation was based on~\cite{tacotron-link}.

In the paper we use the WaveGlow neural vocoder for waveform generation~\cite{glow}. WaveGlow is a deep generative model for audio that integrates Glow, a generative model for image processing,~\cite{NIPS2018_8224} with WaveNet~\cite{wavenet}. We used the WaveGlow implementation at~\cite{glow-link}.

\section{Evaluation}\label{evaluation}
In the section, we evaluated the quality of our four systems objectively and subjectively. We denoted HMM-, DNN-, cGAN-based SPSS systems as HMM, DNN and GAN, respectively. The E2E denotes E2E SPSS system. The NAT denotes target natural speech.  

\begin{table}[b]
\centering
 \begin{tabular}{|c | c | c | c | c|} 
 \hline
  & HMM & DNN & GAN & E2E \\ [0.5ex] 
 \hline\hline
 Decoding & $\textbf{18.9}$ (1.05) & 1.6 (0.47) & 2.4 (0.7) & 5.8 (0.2)  \\ [1ex] 
 \hline
  Vocoding & 5.1 (0.2) & 5.1 (0.2) & 5.1 (0.2) & $\textbf{7.5}$ (0.4)  \\ [1ex] 
 \hline
\end{tabular}
\caption{Performance efficiency in term of RTF. The higher RTF, the better it is. We tested HMM on a single CPU, and other systems (DNN, GAN, E2E) on GPU RTX 2080~Ti). WORLD vocoder was used in HMM, DNN, and GAN; while WaveGlow was used in E2E. }
\label{table:1}
\end{table}

\subsection{Dataset}

All models were trained and tested on ZALO-TTS dataset~\cite{Lam20}. This is a public speech dataset which contains 18 hours of speech from a Vietnamese female speaker. We resampled the waveforms from 44.1 to 22.05~kHz. We randomly selected 32 test sentences. We used the remaining data for training.

\subsection{Training Setup}

For training HMM system, we used the same configurations as in~\cite{Vu09, prosody13}.

For training the DNN system, we used WORLD vocoder to extract 60-dimensional mel-cepstral coefficients (MCCs), five-dimensional band aperiodicities BAPs, and $\log(F0)$ at a 5~ms frame interval. The output features of neural networks thus consisted of MCCs, BAPs, and $\log(F_0)$ with their deltas and delta-deltas, plus a voiced/unvoiced binary feature. The input features were normalised using min-max to the range [0.01, 0.99] and output features were normalised to zero mean and unit variance. The network consisted of six dense layers with 1024 nodes each. A TANH activation function was used. We trained Merlin up to 25~epochs using the a Stochastic Gradient Descent optimizer with a learning rate of $2.10^{-3}$ and a learning rate exponential decay. We used a batch size of 64 and 256 for duration model and acoustic model, respectively.

For training the GAN system, we used the same data presentations as in DNN training. The Generator \emph{G} was a bidirectional recurrent network utilizing Simple Recurrent Units~\cite{lei2018sru} with a dropout rate of 0.2. The Discriminator \emph{D} was a feed-forward network with three layers of 256 nodes each. A TANH activation function was used except for the output layer, which used the sigmoid function. The \emph{G} and \emph{D} components were trained up to 100 and 50 iterations for duration and acoustic modelling, respectively. For duration model training, we used the Adam optimizer~\cite{adam} with a learning rate of $10^{-3}$ for both \emph{G} and \emph{D} components. For acoustic model training, we used the Adagrad optimizer~\cite{JMLR:v12:duchi11a} with a learning rate of $10^{-2}$ for both components. We used a batch size of 32 and 20 for duration model and acoustic model, respectively.

For training the E2E system, we used a Fourier-length of 1024, a hop size of 256 and a window size of 1024. We used a 80-channel mel filterbank spanning from 95~Hz to 7600~Hz. We trained Tacotron-2 up to 50,000 iterations using the a Adam optimizer with a learning rate of $10^{-3}$. We used a decaying rate of $10^{-4}$ from 30,000$^{th}$ iteration to the end. The WaveGlow network was trained up to 350,000 iterations using a weight normalization~\cite{2016arXiv160207868S} and a Adam optimizer with a learning rate of $10^{-4}$. We used a batch size of 32 and 8 for Tacotron-2 and WaveGlow, respectively.

\subsection{Objective Evaluation}

\begin{figure}[t]
\includegraphics[width=0.8\textwidth]{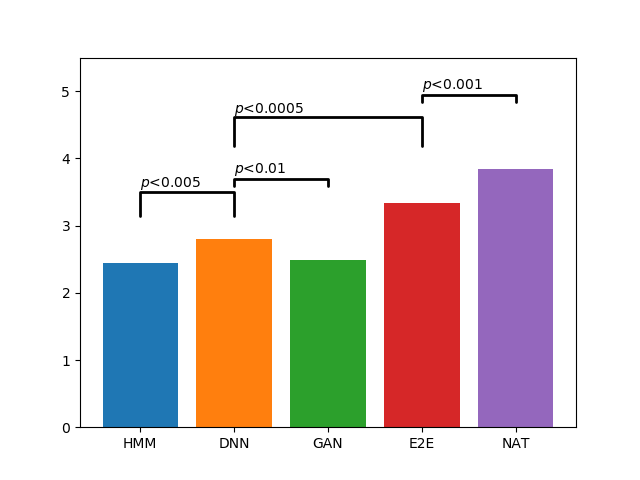}
\caption[width=\columnwidth]{Perceptual MOS scores. The solid lines show statistically significant differences in a two-tailed $t$-test.}\label{fig:MOS results}
\end{figure}

We evaluated the performance efficiency of 1) the decoding process (acoustic modelling, duration modelling and parameter generation), which generates acoustic features from linguistic features, and 2) the vocoding process, which synthesizes waveform from the generated features in term of real time factor (RTF). The RTF tells how many seconds of speech are generated in one second of wall time. A RTF of one means real-time performance. Table~\ref{table:1} shows the average RTF (with standard deviation in parentheses) of four systems. For decoding, the HMM can only run on a single CPU; while DNN, GAN and E2E can utilize the power of GPU. Both DNN, GAN and E2E failed to run on a single CPU. The HMM achieved the fastest RTF of 18.9. The feed-forward network of the DNN was slower than the Tacotron~2 of E2E in term of RTF (1.6 compared to 5.8). For vocoding, both HMM, DNN, and GAN used WORLD vocoder, which ran on CPU; while, E2E used WaveGlow, which required GPU. The WaveGlow (on GPU) was faster than the WORLD vocoder (on CPU) (7.5 compared to 5.1).

\subsection{Subjective Evaluation}

We conducted a mean opinion score (MOS) subjective test. The MOS test consists of 160 (32 sentences $\times$ five systems) unique trials. The perceived loudness differences between these stimuli were minimized using a root-mean-square A-weighted (RMSA) measure~\cite{rmsA}. We need 160 $\div$ 32 = 5 participants to cover all the unique trials. There were 10 participants. For each trial, participants listened to an utterance, and scored the utterance in a five-point scale: "excellent" (5), "good" (4), "fair" (3), "poor" (2), and "bad" (1). The MOS scores of HMM, DNN, GAN, E2E, and NAT are 2.4, 2.8, 2.5, 3.3, and 3.8, respectively (as shown in Figure~\ref{fig:MOS results}). The E2E was the best system, while the HMM was the worst one. Surprisingly, GAN was not as good as DNN. One reason is that training GAN is notoriously difficult, especially on a training data which is not ideal for the task. 

\section{Conclusion}

In the paper, we compared four popular Vietnamese SPSS systems: HMM, DNN, GAN, and E2E in terms of speech quality and performance efficiency. We showed that the E2E systems accomplished the best quality, but required the power of GPU to achieve real-time performance. We also showed that the HMM-based system has inferior speech quality, but it was the most efficient system. We showed that the E2E systems were more efficient than the DNN and GAN in inference on GPU. The GAN did not outperform the DNN in term of quality. Obviously, the HMM is suitable for embedding systems on resource-limited devices (e.g.~a Bose's wireless headphone). While, E2E is suitable for a cloud-based services which utilize the GPU on clusters~(e.g.~Amazon's Alexa).

\bibliographystyle{IEEEtran}  
\bibliography{references}  






\end{document}